\documentclass[aps,preprint,showpacs,amsmath,amssymb, nopreprintnumbers,
floatfix]{revtex4}

\usepackage{graphicx}
\usepackage{dcolumn}
\usepackage{bm}
\usepackage{slashed}
\def\Id{{\rm 1\kern-.3em I}}
\usepackage{natbib,hyperref}
\begin{document}

\title{Nuclear $rho$ meson transparency in a relativistic Glauber model}

\author{W. Cosyn}
\email{Wim.Cosyn@UGent.be}
\author{J. Ryckebusch}

\affiliation{Department of Physics and Astronomy,\\
 Ghent University, Proeftuinstraat 86, B-9000 Gent, Belgium }
\date{\today}

\begin{abstract}
\begin{description}
\item[Background] The recent Jefferson Laboratory data for the nuclear
  transparency in $\rho^ {0}$ electroproduction have the potential to
  settle the scale for the onset of color transparency (CT) in vector
  meson production.
\item[Purpose] To compare the data to calculations in a relativistic
  and quantum-mechanical Glauber model and to investigate whether they
  are in accordance with results including color transparency given
  that the computation of $\rho$-nucleus attenuations is subject to
  some uncertainties.
\item[Method] We compute the nuclear transparencies in a
  multiple-scattering Glauber model and account for effects stemming
  from color transparency, from $\rho$-meson decay, and from
  short-range correlations (SRC) in the final-state interactions (FSI). 
\item[Results] The robustness of the model is tested by comparing the
  mass dependence and the hard-scale dependence of the $A(e,e'p)$
  nuclear transparencies with the data. The hard-scale dependence of
  the $(e,e' \rho ^ {0})$ nuclear transparencies for $^ {12}$C and $^
  {56}$Fe are only moderately affected by SRC and by $\rho^
  {0}$-decay.
\item[Conclusions] The RMSGA calculations confirm the onset of CT at
  four-momentum transfers of a few (GeV/$c$)$^2$ in $\rho$ meson
  electroproduction data.  A more precise determination of the scale for the
onset of CT is hampered by the lack of precise input in the FSI and
$\rho$-meson decay calculations.
\end{description}
\end{abstract}


\maketitle 

\section{Introduction} 
\label{sec:intro}

Color transparency (CT) is a color coherence effect that emerges from
perturbative quantum chromodynamics in exclusive knockout reactions at
high four-momentum transfers $Q^2$ \cite{Brodsky:1982wj,Mueller:1982}.  In
these reactions, a hadron is produced in a small-sized configuration
(SSC) with all quarks sitting close together in the transverse plane.
The color interactions with the surrounding nuclear medium cancel each
other and the hadron can propagate unattenuated as the common
final-state interactions (FSI) between the tagged hadron and the
nuclear environment vanish.  The SSC can also be produced in
nonperturbative conditions \cite{Frankfurt:1992dx,Frankfurt:1993es}.
In this regime, the SSC evolves to its stable hadronic state over
a certain formation length $l_f$.  During its formation the hadron is
subjected to reduced interactions with the nuclear medium.  In order
to observe a CT effect under those conditions, the formation length
should be of the order of the nuclear radius ($l_f \sim R_A$).
Observation of the onset of CT at a certain energy scale can teach us
about the cross-over point between ordinary nuclear matter and
quark-gluon matter.  The identification of this transition point is of
great importance to nucleon-structure studies, as CT is a necessary
condition for the validity of the QCD factorization theorems which are
commonly applied when interpreting data
\cite{Frankfurt:2000jm,Collins:1996fb}.

The measured observable in search for CT is the nuclear transparency
$T$, defined as the ratio of the cross section per target nucleon for
a process on a nucleus to the cross section of the process on a free
nucleon ($T=\sigma^A/A\sigma^N$).  Accordingly, the nuclear
transparency provides a measure of the integrated attenuation of the
nuclear medium on the tagged hadrons in some (semi-)exclusive
reaction. One can study the hard-scale dependence of the transparency
for a certain target nucleus $A$, or the $A$ dependence at a fixed
value of the hard-scale parameter. If CT effects were to appear at a
certain value of the hard-scale parameter, the nuclear transparency
would be observed to overshoot the predictions from traditional
nuclear-physics calculations. The measurement of the onset and
magnitude of the CT effect allows one to constrain models describing
the evolution of a SSC into a hadron.

Experimentally, CT effects have been observed in the measurement of
the cross section of diffractive dissociation of 500 GeV/$c$ pions into
dijets in the E791 experiment at Fermilab \cite{Aitala:2000hc}.  At
intermediate energies, no sign of CT was observed in $A(e,e'p)$
measurements on a variety of nuclear targets and four-momentum
transfers $Q ^{2} \lesssim 8$~(GeV/$c$)$^{2}$
\cite{Garino:1992ca,Makins:1994mm,O'Neill:1994mg,
  Abbott:1997bc,Garrow:2001di,Dutta:2003yt,Rohe:2005vc}.  The nuclear
$^{12}\text{C}(p,2p)$ transparencies were studied at Brookhaven
National Laboratory (BNL)
\cite{Carroll:1988rp,Mardor:1998,Leksanov:2001ui}.  The transparency
first shows a rise with increasing incoming proton momentum and
drops for  momenta larger than 9 GeV/$c$.  This proton-momentum
dependence is at odds with traditional nuclear-physics calculations
predicting $^{12}\text{C}(p,2p)$ nuclear transparencies which are more
or less constant with proton momentum. The BNL $^{12}\text{C}(p,2p)$
results are currently not considered as a clean sign of CT, and
competing effects stemming from nuclear filtering
\cite{Ralston:1988rb,Ralston:1990jj} or from threshold mechanisms for
charm resonance production \cite{Brodsky:1987xw} have been proposed to
explain the observations.

In recent years, several experiments have measured the transparencies
in semi-exclusive meson production reactions.  As a meson is a more
compact object than a baryon, it should be more likely to produce a
meson SSC and observe the onset of CT at intermediate energies.  Two
recent Jefferson Laboratory (JLab) experiments that measured the transparency
of pions in photo- and electroproduction \cite{Dutta:2003mk,
  Clasie:2007gq} consistently agreed with various independent
calculations provided that CT effects are included
\cite{Cosyn:2006vm,Cosyn:2007er,Larson:2006ge,Kaskulov:2008ej}.  More
recently, a JLab Hall-B experiment has measured $\rho^0$ nuclear 
transparencies in semi-exclusive electroproduction on $^{56}$Fe and
$^{12}$C targets \cite{ElFassi:2012nr}.  Again, the data agree
favorably with calculations including CT effects
\cite{Frankfurt:2008pz,Gallmeister:2010wn}.  These results strengthen
the case for an onset of CT at four-momentum transfers of a few
(GeV/$c$)${^2}$ in exclusive meson production reactions.  For a recent
review of the CT phenomenon, see Ref.~\cite{Dutta:2012ii}.

There are issues in nuclear $\rho$ meson transparency calculations, like the
absence of $\rho N$ scattering data, which complicate the
interpretation of the calculations and induce some uncertainties. It
is one of the purposes of this paper to investigate these issues in
more detail and to study the robustness of the computed nuclear $\rho$
transparencies.  We compare the recent $A(e,e'\rho^0)$ transparency
data with calculations in a relativistic multiple-scattering Glauber
approximation (RMSGA) model. The RMSGA model has been used
successfully to predict nuclear transparencies for $A(e,e'p)$
reactions \cite{Ryckebusch:2003fc,Lava:2004zi}, for $A(p,2p)$
reactions \cite{VanOvermeire:2006dr,VanOvermeire:2006tk}, for
$A(\gamma,\pi^- p)$ and $A \left(e,e '\pi ^{+} \right)$ reactions
\cite{Cosyn:2006vm,Cosyn:2007er} and for quasi-elastic
neutrino-induced processes \cite{Martinez:2005xe}.  In
Sec.~\ref{sec:model} we sketch the RMSGA model while highlighting some
important issues emerging for the $\rho^0$ nuclear transparency
calculations.  Numerical results are shown in Sec.~\ref{sec:result}
for kinematics of the JLab experiment. Section~\ref{sec:concl} states our
conclusions.

\section{Model}\label{sec:model}
\paragraph{Relativistic multiple-scattering Glauber approximation}
Finding its roots in optics, Glauber multiple-scattering theory
\cite{Glauber:1959} describes the small-angle scattering of particles
in the eikonal approximation.  Thereby the scattered wave function is
a plane wave multiplied with an eikonal phase.  The eikonal
approximation is valid when the wavelength of the particle is small
in comparison with the typical interaction range of the scattering
particles.  For baryons and mesons this criterion translates into
momenta higher than a few hundred GeV/$c$.  In
Ref.~\cite{Ryckebusch:2003fc} we have introduced a relativistic
version of Glauber multiple-scattering theory.  As the helicity
conserving amplitude is assumed to dominate hadron-nucleon high energy
scattering, the Glauber eikonal phase is a scalar in spin space.  To
describe multiple-rescattering the frozen approximation is used and
the individual eikonal phases are multiplied.  In the RMSGA model, the
eikonal Glauber phase at some spatial point $(\vec{b},z)$ reads for
the ejected $\rho^0$
\begin{equation}
\label{eq:G}
 \mathcal{G}(\vec{b},z)=
 \prod _{\alpha \ne \alpha_{i}} \int d \vec{r}^{\; \prime}
\left| \phi _ {\alpha} \left( \vec{r}^{\; \prime}     \right) \right|^2
\left[
1 -  
  \theta \left( z - z' \right) \Gamma_{\rho N} \left(
\vec{b}^{\; \prime} -
\vec{b} \right) \right]\,.
\end{equation}
Here, the $(\vec{b},z)$ coordinate system has its $z-$axis along the
$\rho^0$ momentum, $\phi_\alpha(\vec{r}\;)$ are the Dirac
single-particle wave functions of the residual nucleons with quantum
numbers $\alpha\equiv n \kappa m_j m_t$ obtained from the Serot-Walecka
model \cite{Furnstahl:1996wv}, and $ \alpha _{i}$ characterizes the
nucleon on which the $\rho ^{0}$ is created.  The profile function
$\Gamma_{\rho N}$ in Eq.~(\ref{eq:G}) is commonly described by a Gaussian
\begin{equation}
\label{eq:gamma}
\Gamma_{\rho N} (\vec{b}) =
\frac{\sigma^{\text{tot}}_{\rho N}(1-i\epsilon_{\rho N})}
{4\pi\beta_{\rho N}^2}\exp{\left(-\frac{\vec{b}^2}{2\beta_{\rho
N}^2}\right)}\,, 
\end{equation}
where $\sigma^{\text{tot}}_{\rho N}, \epsilon_{\rho N}, \beta_{\rho
  N}$ are energy dependent parameters connected to $\rho N$
scattering.  Due to the lack of data for $\rho N$ scattering it has
become customary to use educated estimates for the
$\sigma^{\text{tot}}_{\rho N}, \epsilon_{\rho N}, \beta_{\rho N}$
\cite{Bauer:1977iq} based on the
corresponding values for $\pi N$ scattering. In $\pi N$ scattering the
parameters display very little energy dependence for pion laboratory momenta
$p_\pi \gtrsim
1.5-2~\text{GeV}/c$ and it
has become customary to adopt energy-independent parameters for $\rho N$.
In this work, we study the impact of these uncertainties for the
computed nuclear $\rho$ transparencies.

\begin{figure}
\includegraphics[width=0.5\textwidth]{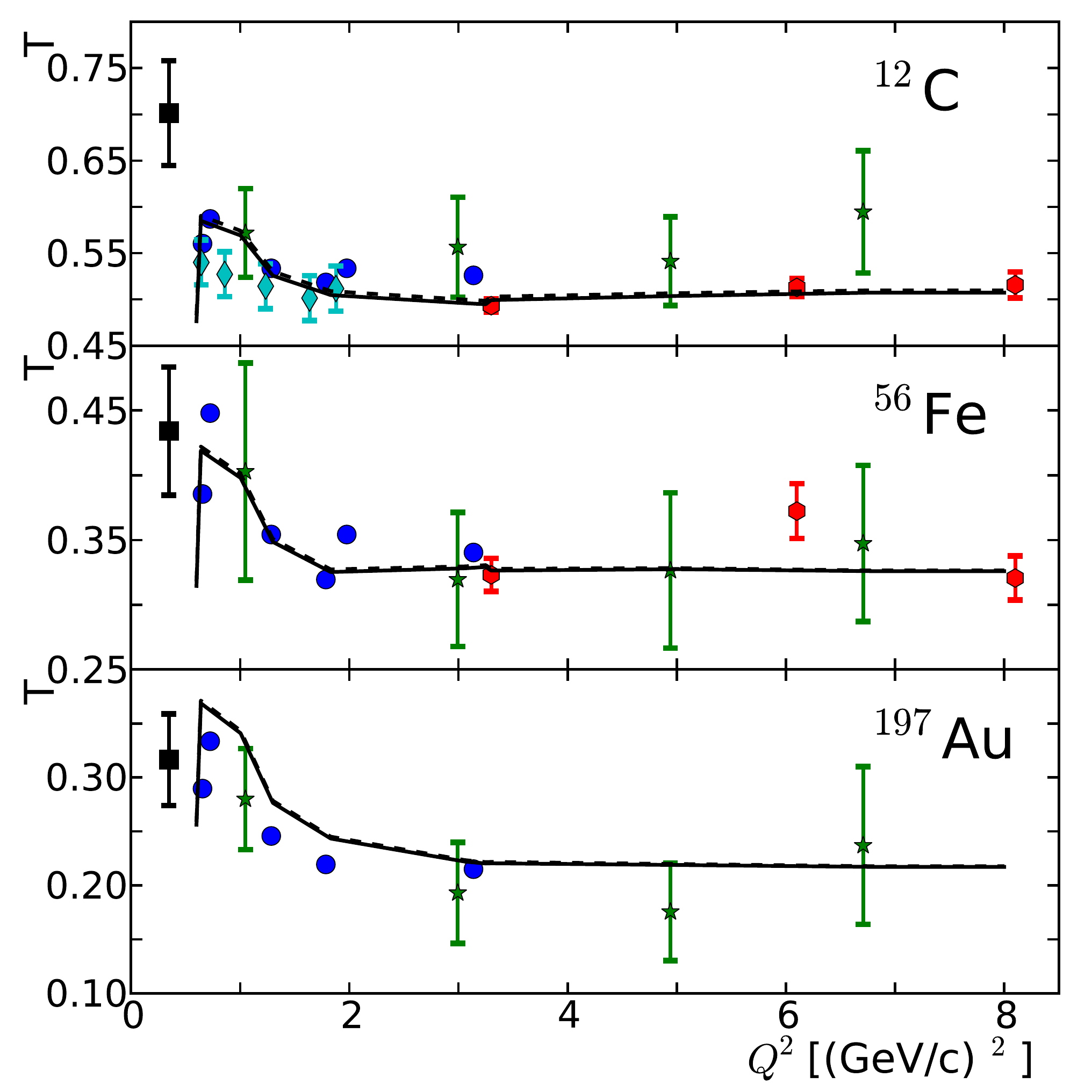}
\caption{(Color online) Nuclear transparencies versus $Q^2$ for
  $A(e,e'p)$ reactions in quasi-elastic kinematics.  Solid black lines are
 RMSGA calculations with the Glauber phase of Eq.~(\ref{eq:G}).  Dashed black
lines
  show the RMSGA calculations corrected for SRC according to the
  replacement of Eq.~(\ref{eq:replaforSRC}) in Eq.~(\ref{eq:G}).  Data
  are from Refs.~\cite{Garino:1992ca} (black squares),
  \cite{O'Neill:1994mg,Makins:1994mm} (red hexagons),
  \cite{Garrow:2001di} (green stars), \cite{Rohe:2005vc} (cyan
  diamonds), and \cite{Abbott:1997bc,Dutta:2003yt} (blue circles).  Data
  and calculations do not include the $c_A$ factor applied in
  \cite{Lava:2004zi}.}.
\label{fig:eepTT}
\end{figure}

\paragraph{Short-range correlations}
The RMSGA model accommodates the possibility to include short-range
correlations (SRC) in the modeling of the FSI.  We use the information
that a nucleon is present at the spatial point of the hard
interaction.  Due to its finite size, the presence of the nucleon
induces local fluctuations in the nuclear density.  The inclusion of
SRC in the FSI is technically achieved in the following way
\cite{Cosyn:2007er}.  First, the squared single-particle wave
functions in Eq. (\ref{eq:G}) can be connected to the one-body density
of the target nucleus $\rho_A ^{[1]}(\vec{r})$ [normalized as $\int
d\vec{r}\, \rho_A ^{[1]}(\vec{r}) = A$]
\begin{equation}
\mid \phi _ {\alpha} ( \vec{r} \; ) \mid ^{2}  \approx 
\frac{\rho_A^{[1]}(\vec{r})}{A}=\int
d\vec{r}_2 \ldots \int d\vec{r}_A
\left(
\Psi_A^{\text{g.s.}}(\vec{r},\vec{r}_2,\ldots,\vec{r}
_A)\right)^\dagger
\Psi_A^{\text{g.s.}}(\vec{r},\vec{r}_2,\ldots,\vec{r}_A)\,.
\end{equation}
Here, $\Psi_A^{\text{g.s.}}$ is the ground-state wave function of the
target nucleus, obtained by antisymmetrizing the product of the
single-particle wave functions $\phi _ {\alpha}$.  Even in a small
nucleus like $^4$He, the above approximation (better known as the thickness
approximation) marginally affects the predicted effect of FSI
\cite{Ryckebusch:2003fc}.  In a second step, the averaged
density $\rho_A ^{[1]}(\vec{r} \; )$ can be substituted with the ratio of
the two-body density $\rho^{[2]}_A$ (normalized as $ \int d\vec{r}_1
\int d \vec{r}_2 \rho^{[2]}_A(\vec{r}_1,\vec{r}_2)=A(A-1)$) and the
one-body density:
\begin{equation}\label{eq:subs}
  \rho^{[1]}_A(\vec{r}\;) \rightarrow
\frac{A}{A-1}\frac{\rho^{[2]}_A(\vec{r},\vec{r}_1)}{\rho^{[1]}_A(\vec{r}_1)}\,,
\end{equation}
where $\vec{r}_1$ is the spatial coordinate corresponding with the
hard interaction.  One can include SRC in the two-body density by
adopting the functional form \cite{Frankel:1992er}:
\begin{equation}
 \rho^{[2]}_{A}(\vec{r}_1,\vec{r}_2)
\approx
\frac{A-1}{A}\gamma(\vec{r}_1)\rho^{[1]}_A(\vec{r}_1)\rho^{[1]}_A(\vec{r}
_2)\gamma(\vec{r}_2) g(r_{12})\,,
\end{equation}
where $g(r_{12})$ is the Jastrow correlation function
\cite{Roth:2010bm} and
$\gamma(\vec{r}\;)$ a function that imposes the normalization of the two-body
density obtained as the solution of an integral equation.  With
the above expression for the SRC-corrected two-body density, 
Eq.~(\ref{eq:subs}) becomes
\begin{equation}
\label{eq:replaforSRC}
 \rho^{[1]}_A(\vec{r}) \rightarrow \gamma(\vec{r}\;) \rho^{[1]}_A(\vec{r}\;)
\gamma(\vec{r}_{1})
g\left(\left|\vec{r}-\vec{r}_1\right|\right) \equiv 
\rho^{\text{eff}}_A(\vec{r}, \vec{r}_1)
\, .
\end{equation}
In summary, the FSI calculations can be corrected for SRC by replacing
$\left| \phi _ {\alpha} \left( \vec{r} \; \right) \right|^2$ with
$\rho^{\text{eff}}_A(\vec{r}, \vec{r}_1)/A$ in Eq.~(\ref{eq:G}).  To
illustrate the effectiveness and robustness of the RMSGA model in
describing a variety of knockout reactions, we show in
Fig.~\ref{fig:eepTT} RMSGA predictions for the $A(e,e'p)$ nuclear
transparencies and compare them to the published data.  Neither the
calculations nor the data include the commonly applied (and rather
arbitrary) correction factors $c_A$ that account for shifts in the
nuclear spectral function to higher missing momenta and energies due
to SRC.  For a discussion on this subject, we refer to
Ref.~\cite{Dutta:2012ii}.  We observe an excellent agreement between
the RMSGA calculations and the world data on $A(e,e'p)$ nuclear
transparencies. Both the hard-scale $Q^{2}$ dependence and the mass
dependence of the data are nicely predicted. Inclusion of the SRC in the FSI
 enhances the $T$ by a tiny amount. The biggest enhancement is of the
order of 0.5\% and is observed for the lowest $Q^2$ values of
the $^{12}$C transparency.  It is noteworthy to
mention that the $A\left(e,e'\rho^{0}\right)$ calculations presented
below use identical nuclear-structure input as used for the $A(e,e'p)$.

\paragraph{Color transparency}
The effects of color transparency are implemented by means of the
quantum diffusion model of Ref.~\cite{Frankfurt:1988nt}.  Thereby, the
position independent parameter $ \sigma^{\text{tot}}_{\rho N} $ in the
profile function of Eq. (\ref{eq:gamma}) is replaced by a
position-dependent effective one $ \sigma^{\text{eff}}_{\rho N} (\mathcal{Z})$
which evolves in a linear fashion along the formation length $l_f$
from a reduced value for the SSC to the standard one associated with
the normal hadron
\begin{equation}
{ \sigma^{\text{eff}}_{\rho N}(\mathcal{Z}) } =   
{ \sigma^{\text{tot}}_{\rho N} } \biggl\{ \biggl[
 \frac{\mathcal{Z}}{l_f} +
 \frac{<n^2 k_t^2>}{\mathcal{H}} \left( 1- \frac{\mathcal{Z}}{l_f} 
\right) \biggr]
\theta(l_f-\mathcal{Z}) + 
\theta(\mathcal{Z}-l_f) \biggr\} \,  \; .
\label{eq:diffusion}
\end{equation}
Here, $n$ is the number of elementary fields ($n=2$ for the $\rho^0$),
$k_t = 0.350~\text{GeV/}c$ is the average transverse momentum of a
quark inside a hadron, $\mathcal{Z}$ is the distance from the hard
interaction point $\vec{r}_{1}$ along the ejected hadron path, and
$\mathcal{H}\equiv Q^2$ is the hard-scale parameter that governs the
CT effect.  Unless otherwise stated, for the formation length $l_f 
\approx 2p/\Delta M^2$, we
adopt the value $\Delta M^2 = 0.7~(\text{GeV}/c^2)^2$. With this value we
could reproduce the measured nuclear pion transparencies
\cite{Clasie:2007gq}. 

\paragraph{$\rho$ decay}
The $\rho^0$ decays to a pair of pions with a branching ratio of
almost 99\%.  For the kinematics of the JLab experiment the average
life time of the $\rho^0$ corresponds to a mean path length of about
5~fm.  This means that the majority of the $\rho^0$ will decay outside
the nuclear medium and the anticipated effect on the computed nuclear
transparencies is rather modest.  The average opening angle of the
pion pair is around 30 degrees in the laboratory frame.  Therefore, it is
fair to substitute the $\sigma^{\text{tot}}_ {\rho N}$ in
Eq.~(\ref{eq:gamma}) by $\left( \sigma^{\text{tot}}_{\pi^-
  N}+\sigma^{\text{tot}}_{\pi^+ N} \right)$ after the decay.  The
$\rho^0$ decay is expected to lower the nuclear transparency as the
pion pair is subject to an increased attenuation compared to the
$\rho^0$.  The adopted procedure gives an upper limit for the effect
of $\rho$ decay as it adds the attenuation on the two pions in an
incoherent way by making them move collinearly \cite{Cosyn:2007er}.  We include
the effect of $\rho^0$
decay by replacing the $\rho N$ total cross section
$\sigma^{\text{tot}}_{\rho N}$ in Eq.~(\ref{eq:gamma}) by a position
dependent one 
\begin{equation}\label{eq:decay}
 \sigma^{\text{tot}}_{\text{decay}}(\mathcal{Z}) = \sigma^{\text{tot}}_{\rho
N}e^{-\mathcal{Z}\Gamma_\rho\sqrt{1-|\vec{p}_\rho|^2/E_\rho^2}} +
(\sigma^{\text{tot}}_{\pi^-
N}+\sigma^{\text{tot}}_{\pi^+
N})(1-e^{-\mathcal{Z}\Gamma_\rho\sqrt{1-|\vec{p}_\rho|^2/E_\rho^2}})\,.
\end{equation}
Here, $\vec{p}_\rho$ ($E_\rho$) are the rho meson momentum (energy),
and $\Gamma_\rho=149$ MeV is the $\rho^0$ decay width in the laboratory
frame.

\paragraph{Cross section}
The cross section for $\rho^0$ electroproduction on a nucleus $A$ ($e+A
\rightarrow e' + \rho^0 + A^*$) takes the following form (Dirac spinors are
normalized as $\bar{u}u=1$)
\begin{equation}\label{eq:A}
 \frac{d\sigma^{eA}}{dE_{e'} d\Omega_{e'} dt d\phi_{\rho}} =\sum_{\alpha}
\frac{\alpha_{\text{EM}}}{8(2\pi)^4}\frac{E_{e'}}{q
E_eQ^2(1-\epsilon)
}\overline{\sum}|\mathcal{M}^{\gamma^*A}_\alpha|^2\,.
\end{equation}
Here, the summation runs over all the shells $\alpha$ of the nucleus
$A$, $E_e$ and $E_{e'}$ are incoming and scattered electron energy,
$\alpha_{\text{EM}}$ is the fine structure constant, the virtual
photon $q^\mu(\nu,\vec{q})$ has four-momentum transfer $Q^2 = -q^2$
and degree of transverse polarization $\epsilon$, $t=(q-p_\rho)^2$,
$\overline{\sum}$ denotes the summing and averaging over spin degrees
of freedom, and $\mathcal{M}^{\gamma^*A}$ is the matrix element of the
$\gamma^*+A \rightarrow \rho^0 + A^*$ reaction (with $A^*$ the residual
nucleus) for $\gamma^*$ absorption on a nucleon with quantum numbers $\alpha$. 
In a factorized approach, the squared matrix element can be
approximated as \cite{Cosyn:2007er,VanOvermeire:2006dr}
\begin{equation}\label{eq:factorize}
 \overline{\sum}|\mathcal{M}^{\gamma^*A}|^2 \approx \int d\vec{p}_m
\overline{\sum}|\mathcal{M}^{\gamma^*N}|^2 \rho^D_\alpha(\vec{p}_m)\,,
\end{equation}
where $\mathcal{M}^{\gamma^*N}$ is the amplitude of the $\gamma^*+N
\rightarrow \rho^0 + N$ process, $\vec{p}_m = \vec{p}_{\rho} -\vec{q}$
is the missing momentum, and the distorted momentum distribution
$\rho^D_\alpha$ is defined as
\begin{equation}\label{eq:distormomentum}
 \rho^D_\alpha(\vec{p}_m) =\frac{1}{(2\pi)^3}
\sum_{m_j,m_s}\left|\int d\vec{r}
e^{-i\vec{p}_m\cdot\vec{r}}\mathcal{G}(\vec{b},z)\bar{u}(\vec{p}_m,
m_s)\phi_\alpha(\vec{r})\right|^2\,.
\end{equation}
In Eq.(\ref{eq:factorize}), the matrix element can be related to the
cross section in the center-of-mass frame:
\begin{equation}\label{eq:elem}
 \frac{d\sigma^{\gamma^*N}}{d|t|d\phi^*} =
\frac{m_N^2}{8\pi^2(s^2-2s(m_N^2-Q^2)+(m_N^2+Q^2)^2)}\overline{\sum}\mathcal{M}^
{\gamma^*N} \approx \left(\frac{d\sigma}{d|t|}\right)_0
e^{-\beta_{\gamma\rho} t}\,,
\end{equation}
where $s=W ^{2}$, the squared c.m. energy for the $\gamma^* N$ system.
In the last step we made use of the diffractive properties of the
vector meson cross section at GeV energies.  For the slope
factor of the diffractive $\rho^0$ production, we take $\beta _{\gamma\rho
}=6$~(GeV/$c$)$^{-2}$
\cite{Bauer:1977iq,Arneodo:1994id,Derrick:1995yd,Aid:1996ee}  and leave
$\left(\frac{d\sigma}{d|t|}\right)_0$ unspecified as it cancels in the
nuclear transparency ratio $T=\sigma^A/A\sigma^N$. Combining Eqs.~(\ref{eq:A}),
(\ref{eq:factorize}) and (\ref{eq:elem}) gives us the following
formula for the cross section:
\begin{multline} \label{eq:cross}
  \frac{d\sigma^{e A}}{dE_{e'} d\Omega_{e'} dt d\phi_{\rho}} =\sum_{\alpha} \int
d\vec{p}_m
\frac{\alpha_{\text{EM}}}{4(2\pi)^2}\\ \times \frac{E_{e'}\rho^D_\alpha(\vec{
p}_m) \left[s^2-2s(m_N^2-Q^2)+(m_N^2+Q^2)^2\right]}{q
E_eQ^2(1-\epsilon)m_N^2
}\left(\frac{d\sigma}{d|t|}\right)_0e^{-\beta _{\gamma\rho} t}\,.
\end{multline}

\section{Results}
\label{sec:result}
The JLab Hall-B experiment E02-110 measured the nuclear transparency
in $\rho^0$ electroproduction for $ 0.8 \le Q^2 \le 3$~(GeV/$c$)$^2$ on
$^{12}$C and $^{56}$Fe targets using the CLAS $4\pi$ detector.  In
order to select elastically produced $\rho^0$ mesons and suppress
pions from resonance decays the following kinematical cuts were
imposed: $z(=E_\rho/\nu) > 0.9$ and $W>2$~GeV. To ensure the selection
of exclusive diffractive and incoherent events, $t$ was limited to
$-0.4<t<-0.1$ (GeV/$c$)$^2$.  In vector meson electroproduction, the
virtual photon will fluctuate into a $q\bar{q}$ pair along a certain
coherence length $l_c=2\nu/(Q^2+m^2_{q\bar{q}})$ and then scatter off
a nucleon.  To isolate a possible CT signal it is essential that $l_c$
is more or less constant over the kinematic ranges included in the
analysis.  As the $q\bar{q}$ is subject to initial-state interactions
(ISI), a variation in $l_c$ can cause a change in the transparency and
thus mimic a CT signal.  For the JLab experiment E02-110 one has $0.5 \le l_c
\le 0.8$~fm.  This is sufficiently smaller than the typical
intra-nucleon distance in the nucleus, so ISI are not included in our
calculations.

The nuclear transparencies shown here are computed as the ratio
of the cross section of Eq.~(\ref{eq:cross}) to its plane-wave
approximation (PWA) equivalent.  The PWA cross section neglects any
form of FSI and is obtained by putting $\mathcal{G}= 1$ in
Eq.~(\ref{eq:distormomentum}).  It deserves highlighting that the
experimentally extracted nuclear $\rho$ transparencies use a ratio of
the phase-space integrated cross sections from nucleus $A$ to the
deuteron.  In Ref.~\cite{Gallmeister:2010wn} it is pointed out that
Fermi motion can introduce an overall change in the transparency
with a constant factor and mimic
part of a CT signal through a $Q^2$ dependent change of the plane-wave ratio. 
The results of
Ref.~\cite{Gallmeister:2010wn} suggest that in the PWA the ratio of
the cross section on $A$ relative to the deuteron increases with
$Q^{2}$ due to the applied cuts in $t$ and the effect of Fermi motion.
For the calculations presented here we prefer to compute the
transparencies as a ratio of the RMSGA to the PWA cross sections for
$^{12}$C and $^{56}$Fe. Indeed, the nuclear transparency computed as a
RMSGA/PWA ratio for the same target nucleus is less prone to errors
stemming from specific phase-space mismatches between $^{12}$C,
$^{56}$Fe and the deuteron for example, as well as other uncertainties
as possible medium-effects on the $\gamma^*+N \rightarrow \rho + N$
reaction amplitude.

The experimental kinematical cuts determine the phase space of the
calculations.  For each $Q^2$ bin, we fix $Q^2$ at its central value
and compute the RMSGA and PWA cross sections for the ranges in $(\nu,
z, t)$ used in the analysis of the data. The final result for the
nuclear transparency at a particular value of the hard-scale parameter
$Q^2$ is computed as the ratio of the integrated RMSGA cross section
to the integrated PWA one, whereby both in the nominator and the denominator
the same ranges in  $(\nu, z, t)$ are considered.

\begin{figure}
\includegraphics[width=0.7\textwidth]{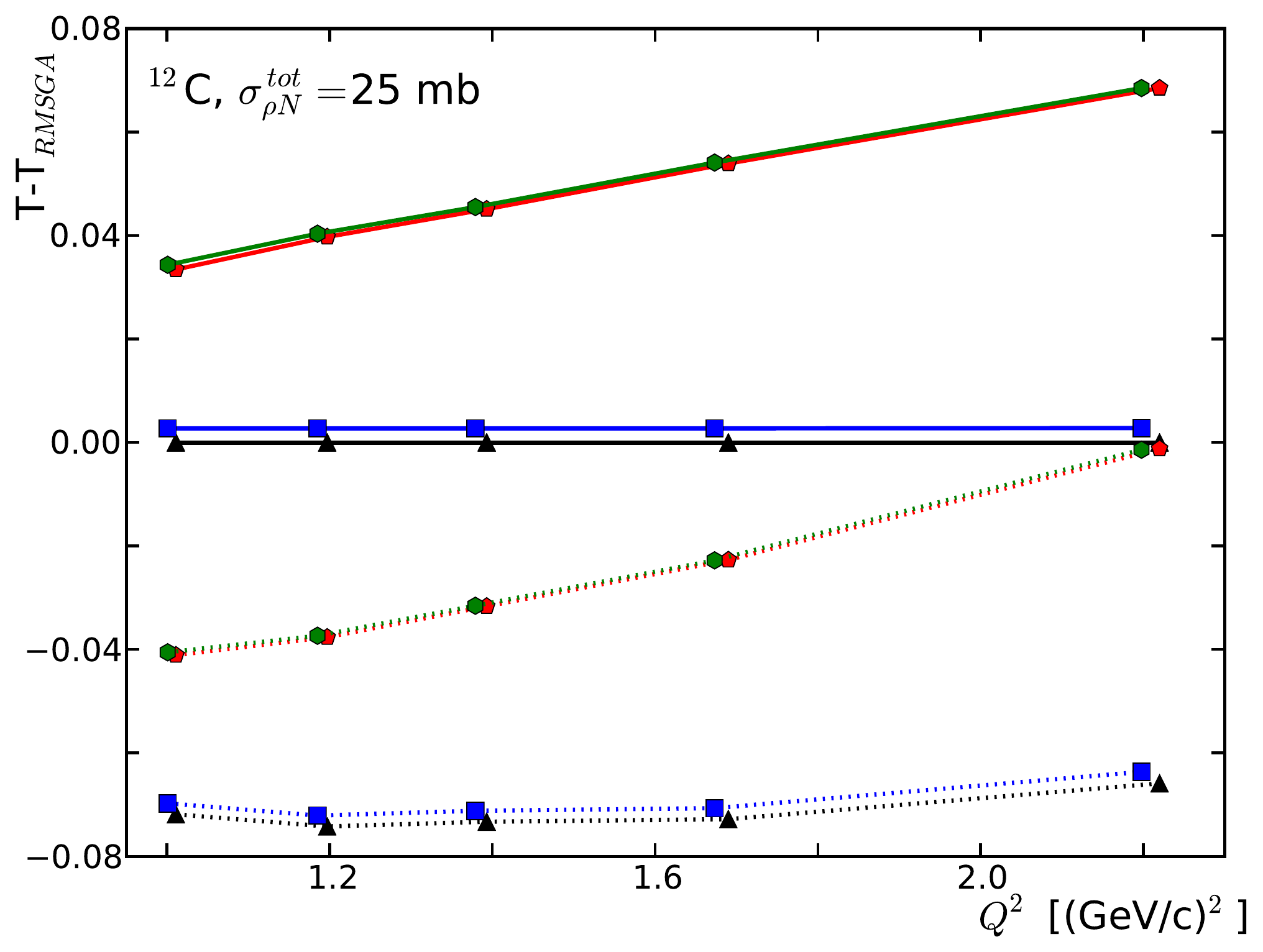}
\caption{(Color online) Relative effect of the SRC, of the
  $\rho$-decay, and of the CT on the predicted hard-scale dependence
  of the $^{12}$C$(e,e' \rho ^ {0})$ nuclear transparencies. All
  calculations use $\sigma _{\rho N} ^ {\text{tot}}$=25~mb and $\beta _{\rho
    N}=6~(\text{GeV}/c)^{-2}$. Base-line calculations (denoted as
  $T_{\text{RMSGA}}$) are obtained with Eqs.~(\ref{eq:G}) and
  (\ref{eq:gamma}). Dashed (solid) curves include (exclude) the effect
  of $\rho$ decay implemented with the Eq.~(\ref{eq:decay}).  Black
  triangles (blue squares) exclude (include) the effect of SRC in the
  computation of the FSI. Green pentagrams include the effect of CT
  and red hexagons include both SRC and CT. Some points are offset on the
$x$-axis for clarity. The kinematic cuts are
  those of the JLab E02-110 experiment (details in text).}
\label{fig:RMSGACbis}
\end{figure}

Figure \ref{fig:RMSGACbis} shows the hard-scale dependence of the
RMSGA nuclear $^{12}$C$(e,e ' \rho ^{0})$ transparencies for
$\sigma^{\text{tot}}_{\rho N}$=25~mb, $\beta_{\rho N}=
6~(\text{GeV/$c$})^{-2}$, $\epsilon_{\rho N}=-0.2$ taken constant as a
function of energy. We investigate the hard-scale dependence stemming
from the effect of CT, $\rho$ decay, and SRC.  The computed
transparencies that do not include CT show little dependence on $Q^2$
as can be anticipated from the energy independence of the Glauber
scattering parameters at these energies.  The SRC component in the FSI
adds about 0.5\% to the transparency, independent of the value of the
hard scale parameter.  Accordingly, the CT and SRC mechanisms can be
separated by studying the hard-scale dependence of the
transparency. The $\rho$ decay lowers the transparency with about 7\%
at the lowest $Q^2$  data point and 6\% at the highest $Q^2$ data point,
reflecting the longer dilated $\rho$ half-life time at higher
energies.  The CT effects induce the strongest $Q^2$ dependence,
increasing the transparency with 3 to 6\% from low to high $Q^2$.  The
inclusion of SRC in the FSI for the CT calculations yields almost no
increase of the transparency as the produced $\rho^0$ in the SSC is
already subjected to reduced interactions in the local neighborhood of
the hard interaction where also the SRC effectively modify the
density.

\begin{figure}
\includegraphics[width=0.7\textwidth]{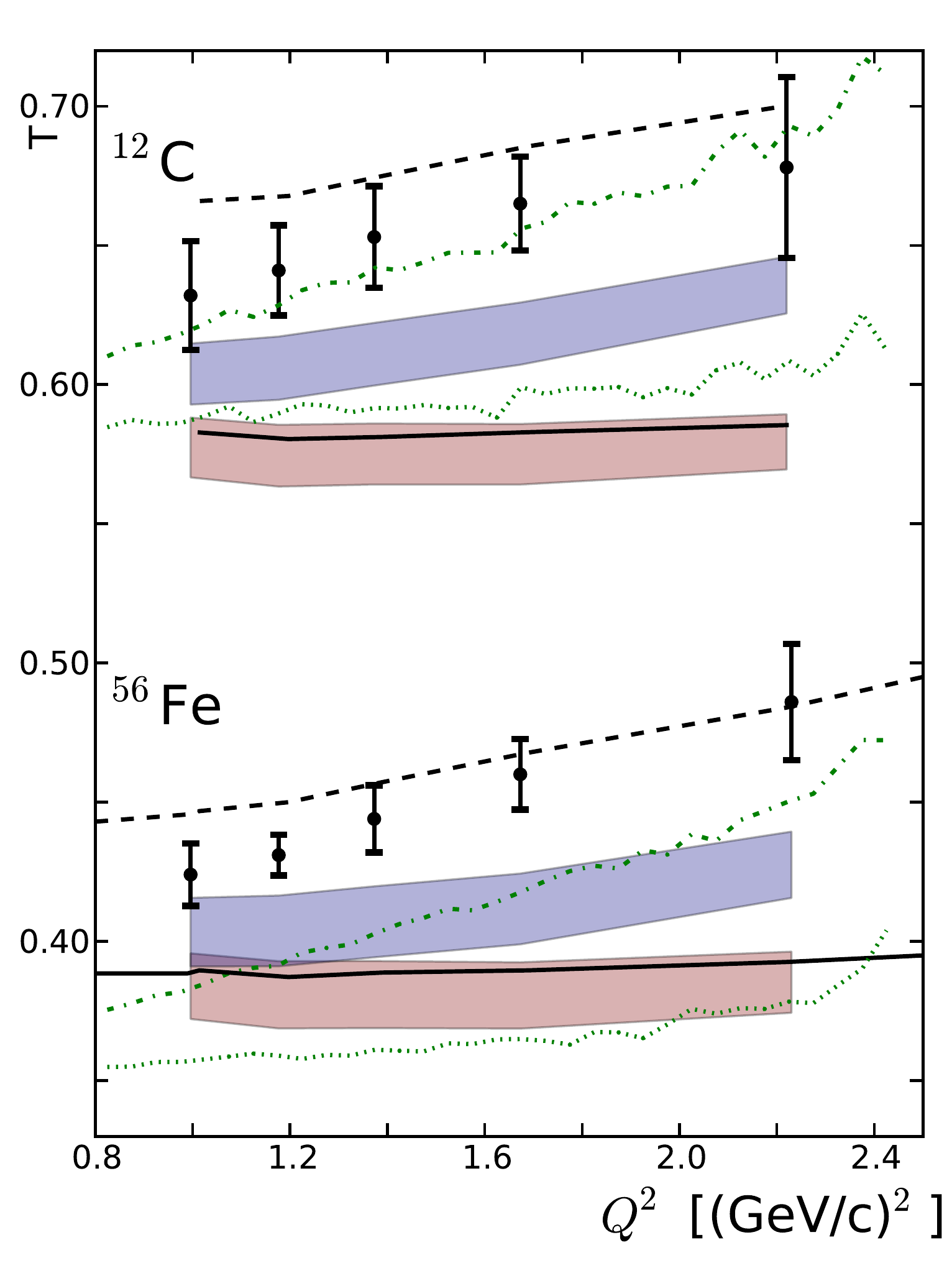}
\caption{(Color online) Nuclear transparency for $\rho$
  electroproduction as a function of $Q^2$ in $^{12}$C and $^{56}$Fe
  with $\sigma^{\text{tot}}_{\rho N}=20$~mb.  Data are from
\cite{ElFassi:2012nr}.
  Black curves are calculations from the model of
  Ref.~\cite{Frankfurt:2008pz} with (dashed) and without (full) CT
  effects.  Green curves are calculations from the model of
  Ref.~\cite{Gallmeister:2010wn} with (dash-dotted) and without
  (dotted) CT effects.  Blue (with CT effects) and red (no CT effects)
  shaded bands are results from the RMSGA model including rho decay
  and SRC effects.}
\label{fig:data20}
\end{figure}

\begin{figure}
\includegraphics[width=0.7\textwidth]{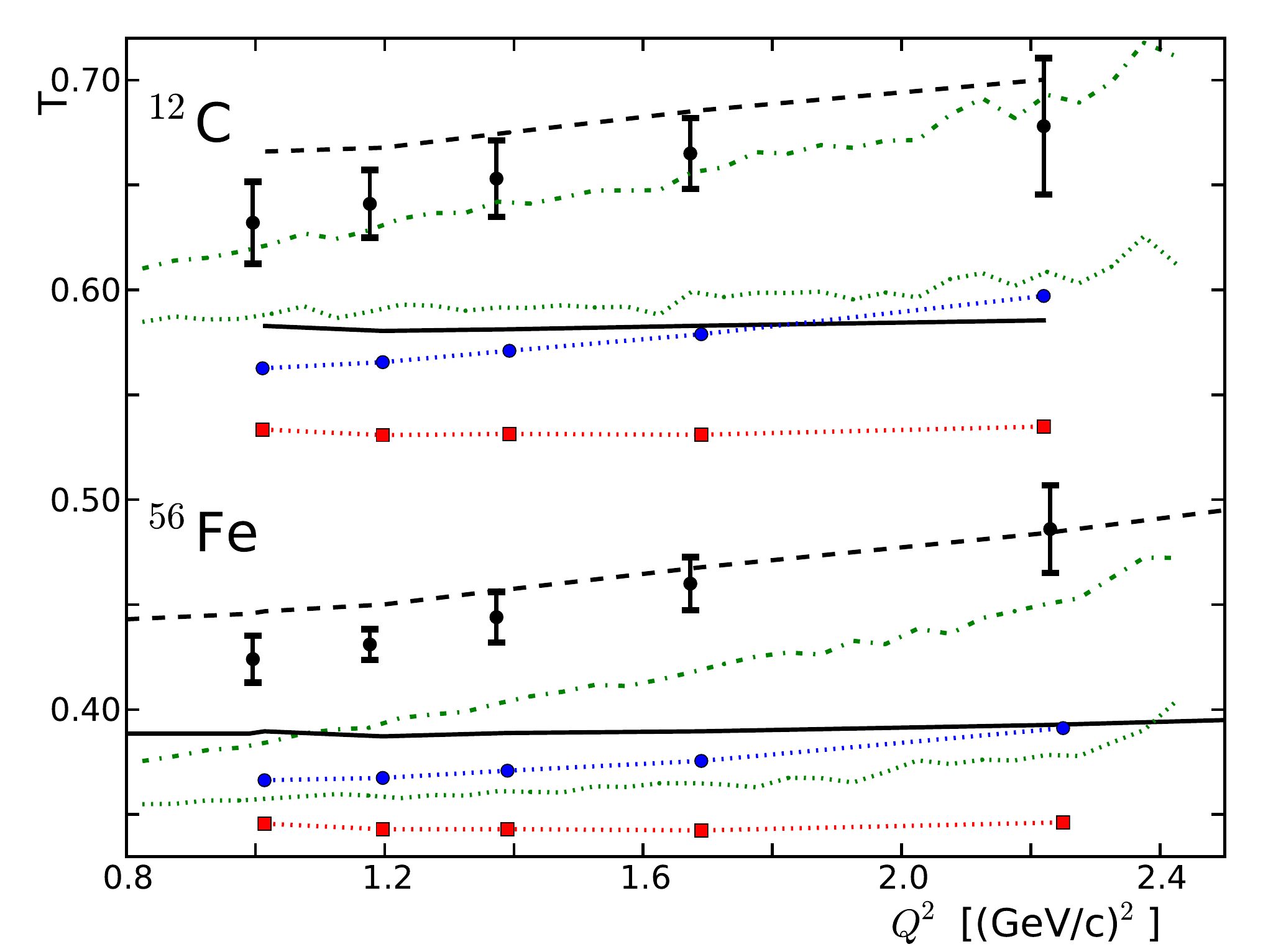}
\caption{(Color online) As in Fig.~\ref{fig:data20} but for
$\sigma^{\text{tot}}_{\rho
N}=25$ mb. Blue circles (with CT effects) and red squares (no CT effects)
  curves are results from the RMSGA model including rho decay
  and SRC effects.}
\label{fig:data25}
\end{figure}

Figures \ref{fig:data20} and \ref{fig:data25} compare the RMSGA
transparencies including SRC and $\rho$ decay with the JLab data.
Given the lack of detailed information on the $\sigma_{\rho
  N}^{\text{tot}}$ parameter, we show results of calculations for
$\sigma_{\rho N}^{\text{tot}}=20$~mb (Fig.~\ref{fig:data20}), and for
$\sigma_{\rho N}^{\text{tot}}=25$~mb (Fig. \ref{fig:data25}). As
mentioned in Sec.~\ref{sec:model} the computed transparencies including 
the decay
of the $\rho$ meson by means of Eq.~(\ref{eq:decay}) 
represent a lower limit.  We therefore have chosen
to represent our calculations in Fig. \ref{fig:data20} as a shaded region,
confined on the
lower side by the calculations with the $\rho$ meson decay evaluated
with the expression for the $\rho N$ total cross section of
Eq.~(\ref{eq:decay}) which assumes that the two pions move collinearly.  
We have also evaluated the
transparency of two pions decaying from a 3~GeV $\rho$ meson 
(representative for
the kinematics of the data) with various $\pi-\rho$ c.o.m. angles and have  
compared this to value of the transparency for two collinear pions. We find  
a maximum enhancement 
of the two-pion transparency of 28\% for $^{12}$C and 20\% for $^{56}$Fe
 compared to the collinear situation.  The upper limits of the bands in 
Fig.~\ref{fig:data20} are obtained by reducing the effect of the inclusion 
of rho meson decay by  
28\% for $^{12}$C and 20\% for $^{56}$Fe. This is a fair estimate 
of the maximum impact of the incoherence effect. Therefore, the width of the band
reflects the estimated uncertainty in the computation of the effect of
the $\rho^{0}$ decay.  In Fig.~\ref{fig:data25} we only show the curves 
corresponding with the lower limit of the bands in order to compare 
to alternate model calculations.

Figures \ref{fig:data20} and \ref{fig:data25} also include two
other model calculations. The model by Frankfurt, Miller, and Strikman
(FMS) \cite{Frankfurt:2008pz} is based on a semi-classical Glauber
calculation and implements the effects of CT and $\rho$ decay along
similar lines as ours, with the values of $\sigma_{\rho
  N}^{\text{tot}}$=25~mb, $\Delta M^2=0.7~(\text{GeV}/c^2)^2$, and
$\Gamma_\rho=149$ MeV. The calculations of
Ref.~\cite{Gallmeister:2010wn} from the Giessen group include a model
for the elementary $\rho$ production and describe the FSI with the
semi-classical GiBUU transport model \cite{Buss20121,GiBUU}.  Given
their very different nature, it is satisfying that all three models
yield similar $^{12}$C transparencies when ignoring CT.
The predictions including CT effects show a
little more variation over the different groups, but display similar
trends in their $Q^2$ dependence.  At identical parameter input in the Glauber
part, we obtain a transparency that is about 5\% lower than those of the FMS model.  The difference between the calculations including and excluding
CT effects is also bigger in the FMS model.  It is worth noting that when
considering the
$A$-dependence, none of the models can satisfyingly describe the data for both
nuclei with
the same parameter set.  This was not the case for the $A(e,e'p)$ and
$A(e,e'\pi^+)$ data, where one parameter set gave a very good agreement over
the whole measured $A$-range.  The RMSGA
results with $\sigma_{\rho
  N}^{\text{tot}}=20$~mb are a better match for the data than those
with $\sigma_{\rho N}^{\text{tot}}=25$~mb, but are underestimating both
the magnitude and the $Q^2$ slope of the data.

\begin{figure}
\includegraphics[width=0.7\textwidth]{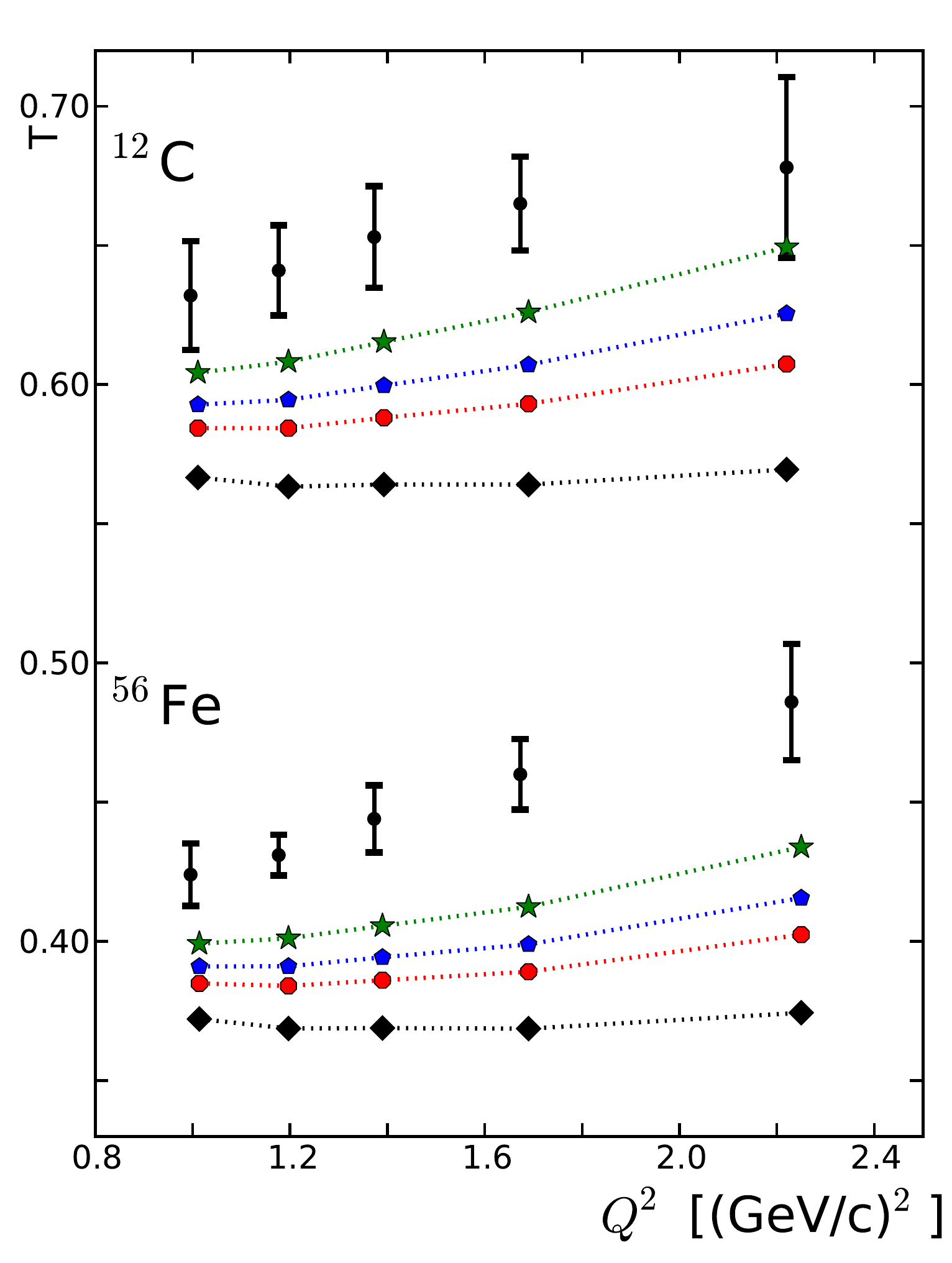}
\caption{(Color online) The hard-scale dependence of the nuclear 
rho transparencies for $\sigma^{\text{tot}}_{\rho N}=20$~mb and 
various choices of the formation length: 
$\Delta M^2=0.5~(\text{GeV}/c^2)^2$ (green stars), $\Delta M^2=0.7~(\text{GeV}/c^2)^2$ (blue
pentagrams) and $\Delta M^2=1.0~(\text{GeV}/c^2)^2$ (red circles).  The black diamonds curve does not
include CT effects.  All curves include the $\rho$-meson decay and SRC
effects.}
\label{fig:dataCT}
\end{figure}

In the quantum diffusion CT model of Eq.~(\ref{eq:diffusion}) the CT
effect can be made bigger by decreasing the value of $\Delta M^2$. Up
to this point, $\Delta M^2=0.7~(\text{GeV}/c^2)^2$ has been used. To
date, however, there is little guidance with regard to realistic
ranges for those parameters.  In Fig.~\ref{fig:dataCT} we show
calculations for three different $\Delta M^2$ values.  The $\Delta
M^2=1~(\text{GeV}/c^2)^2$ corresponds with the value commonly adopted
for the nucleon.  It is clear that the results with $\Delta
M^2=0.5~(\text{GeV}/c^2)^2$ yield the best correspondence with the
data, both with regard to the magnitudes and $Q^{2}$ slope.
Any further fine tuning of the parameters is not opportune given the mentioned
uncertainties in the calculations. 

In Table \ref{tab:slopes} we assess the computed and measured slopes
of the hard-scale dependence of the rho transparencies.  We clearly
see a better agreement between the measured slopes and those
calculations including CT.  As becomes clear from
Fig.~\ref{fig:data20} the FMS and RMSGA calculations with $\Delta
M^2=0.7~(\text{GeV}/c^2)^2$ tend to underestimate the measured slopes.
While the computed slope with $\Delta M^2=0.5~(\text{GeV}/c^2)^2$ is
within the error bars for $^{12}$C, it is substantially
underestimating the measured $^{56}$Fe one.  The GiBUU predictions show a
stronger $Q^{2}$ dependence than our calculations. This can be
explained by the effect of Fermi motion in large nuclei mentioned
earlier in this section, which induces an additional rise in $T$ with
$Q^2$ on top of the one due to the CT effect.

\begin{table}[htb]\label{tab:slopes}
  \centering
  \begin{tabular}{|l|c|c|c|c|c|c|c|}
    \hline 
    nucleus & $\sigma_{\rho N}^{\text{tot}}$& $\Delta M^2$&JLab data
\cite{ElFassi:2012nr} &
RMSGA (+SRC & RMSGA &
RMSGA & RMSGA \\    
& [mb]& $(\text{GeV}/c^2)^2$ & &+CT+decay)& (+SRC+CT)& (+SRC+decay) & (+SRC)  \\
    \hline
\hline
$^{12}$C & 25 & 0.5& 0.044$\pm$ 0.015 $\pm$ 0.019 & 0.040 & 0.034
&0.0017&-0.0038\\
  $^{12}$C & 25 & 0.7& & 0.029 & 0.024&0.0017&-0.0038 \\
  $^{12}$C & 25 & 1.0& & 0.020 & 0.016&0.0017&-0.0038 \\
 $^{12}$C & 20 & 0.5&0.044$\pm$ 0.015 $\pm$ 0.019 & 0.038 & 0.030
&0.0032&-0.0037\\
 $^{12}$C & 20 & 0.7& & 0.028 & 0.021&0.0032&-0.0037 \\
 $^{12}$C & 20 & 1.0& & 0.020 & 0.014 &0.0032&-0.0037 \\
  $^{56}$Fe & 25 & 0.5&0.053 $\pm$ 0.008$\pm$ 0.013 & 0.029 & 0.026
&0.0011&-0.0050 \\
  $^{56}$Fe & 25 & 0.7& & 0.020 & 0.017&0.0011&-0.0050 \\
  $^{56}$Fe & 25 & 1.0& & 0.014 & 0.011&0.0011&-0.0050 \\
  $^{56}$Fe & 20 & 0.5&0.053 $\pm$ 0.008$\pm$ 0.013 & 0.029 & 0.025 
& 0.0026 &-0.0050\\ 
  $^{56}$Fe & 20 & 0.7& & 0.021 & 0.016 & 0.0026 &-0.0050 \\ 
  $^{56}$Fe & 20 & 1.0&  & 0.015 & 0.010  & 0.0026 &-0.0050 \\ 
    \hline
  \end{tabular}
\caption {Slopes for the hard scale dependence of the
$\rho$ nuclear transparencies. Different RMSGA calculations are 
compared to the JLab data.}
\end{table}

\section{Conclusion} \label{sec:concl} We have performed relativistic
Glauber calculations for $A(e,e' \rho ^{0} )$ transparencies and
compared them to the recent JLab data.  The effects of short-range
correlations, color transparency, and the decay of the $\rho$ meson to
a pair of pions can be included.  The covered phase space in the
calculations matches the experimental conditions and all kinematical
cuts are taken into account.  The predicted effect of SRC in the
final-state interactions on the computed transparencies is small, or
even negligible after also including CT effects.  Including the $\rho$
meson decay lowers the transparency up to 6-7 \%, with a smaller
decrease for higher $Q^2$.  This reflects the higher absorption rate
of the two pions compared to the $\rho$.

When comparing to the data and other model calculations, the results
including the CT effect are consistently in better agreement than
those without.  The data suggest a stronger hard-scale dependence of
the transparencies than predicted by our calculations using educated
estimates for the parameters determining the magnitude of the CT
effect.  The presented comparison between calculations and data, gives
additional support for the onset of CT in meson electroproduction
reactions at energies of a few (GeV$/c)^2$.

\subsection*{ACKNOWLEDGMENTS}

The authors are thankful to Lamiaa El Fassi for useful discussions.
The computational resources (Stevin Supercomputer Infrastructure) and
services used in this work were provided by Ghent University, the
Hercules Foundation and the Flemish Government – department EWI.  This
work is supported by the Research Foundation Flanders.

\bibliography{../../bibtexall.bib}
\end{document}